\documentclass[%
 reprint,
 amsmath,amssymb,aps,prl,
]{revtex4-1}

\usepackage{graphicx}% Include figure files
\usepackage{dcolumn}% Align table columns on decimal point
\usepackage{bm}% bold math
\newcommand{\vecE}{{\bf{E}}}
\newcommand{\vecB}{{\bf{B}}}

\newcommand{\phib}{\left<\phi\right>}
\newcommand{\nap}{\nabla_{\perp}}
\newcommand{\dy}{\partial_y}
\newcommand{\dx}{\partial_x}
\newcommand{\dz}{\partial_z}
\newcommand{\dt}{\partial_t}
\newcommand{\curv}{\hat{C}}
\newcommand{\dpar}{\partial_{\parallel}}
\newcommand{\vpar}{\upsilon_{\parallel}}
\newcommand{\vy}{\upsilon_{E_y}}
\newcommand{\vyb}{\left<\vy\right>}
\newcommand{\avg}[1]{\left\langle #1 \right\rangle}
\newcommand{\vx}{\upsilon_{E_x}}
\newcommand{\vdix}{\upsilon_{dix}}

\newcommand{\Rt}{R_t}
\newcommand{\Rp}{R_{\perp}}
\newcommand{\Rpar}{R_{\parallel}}
\newcommand{\Nk}{N_{\boldsymbol{k}}}
\newcommand{\Nkz}{N_{\boldsymbol{k},0}}
\newcommand{\nax}{\nabla_{\boldsymbol x}}
\newcommand{\nak}{\nabla_{\boldsymbol k}}
\newcommand{\kx}{k_x}
\newcommand{\dkx}{\partial_{k_x}}
\newcommand{\ky}{k_y}
\newcommand{\kp}{k_{\perp}}

\newcommand{\dtk}{\mathnormal{d}k^2}
\newcommand{\cop}{\Delta\omega}
\newcommand{\Kx}{K_x}
\newcommand{\Ky}{K_y}

\newcommand{\vgx}{v_{gx}}

\newcommand{\odw}{\omega_0}
\newcommand{\al}{\alpha}
\newcommand{\tal}{\tilde{\alpha}}
\newcommand{\pik}{\pi_{\boldsymbol{k}}}

\newcommand{\avx}[1]{\left<#1\right>_x}
\newcommand{\avyz}[1]{\left<#1\right>}
\newcommand{\avk}[1]{\left<#1\right>_k}

\newcommand{\nfac}{n(\boldsymbol{k})}
\newcommand{\meanu}{\avx{u^2}}
\newcommand{\kl}{K_{x,l}}
\newcommand{\kh}{K_{x,h}}
\newcommand{\dq}{\dx\ln Q}
\newcommand{\alzp}{\al_{0,\perp}}
\newcommand{\ri}{\rho_i}

\newcommand{\lrp}[1]{\left(#1\right)}
\newcommand{\epsn}{\epsilon_n}

\begin{document}
\preprint{PRL}

\title{Predicting Zonal Flows -- A Comprehensive Reynolds-Stress Response-Functional from First-Principles-Plasma-Turbulence Computations}% Force line breaks with \\

\author{Niels Guertler}
 \email{niels.guertler@ipp.mpg.de}
\author{Klaus Hallatschek}%
\affiliation{%
 Max-Planck-Institut f\"ur Plasmaphysik\\
85748 Garching, Germany
}%

\date{\today}% It is always \today, today,

\begin{abstract}
Turbulence driven zonal flows play an important role in fusion devices since they improve plasma confinement by limiting the level of anomalous transport. Current theories mostly focus on flow excitation but do not self-consistently describe the nearly stationary zonal flow turbulence equilibrium state. First-principles two-fluid turbulence studies are used to construct a Reynolds stress response functional from observations in turbulent states. This permits, for the first time, a reliable charting of zonal flow turbulence equilibria.
\end{abstract}

\maketitle

\paragraph{\label{intro}Introduction.---}
Zonal flows (ZF) in toroidal fusion devices are flux-surface averages of layered radial electric fields causing poloidal $\vecE\times\vecB$ flows with zero poloidal and toroidal mode numbers. Stationary ZFs, dominant in the core region, are governed by Reynolds stress (RS) and reduce the level of anomalous transport by ion-temperature-gradient (ITG) turbulence by orders of magnitude \cite{linhahm_zfshear,rosenbHinton_zfshear,hammett_zfshear}. Hence, it is imperative to understand the ZF evolution and take their influence into account for anomalous transport predictions. Nonlinear analytic ZF theories are largely based on wave-kinetics \cite{hasegawa_zfgen,diamondKim_zfgen,diamondRosenbluth_IAEA_zfgen,itoh_zfphysics,itoh_hallatschek_zfwavek,diamondHallatschek_instabdrift,Itoh2005} and wave-kinetic effects have been numerically observed in \cite{gamprl}. But most numerical studies are restricted to the observation of exponential ZF growth and anomalous transport reduction \cite{dorland_zfinstab,linhahm_zfinstab}. In order to understand the ZF evolution and the radial scale length observed in experiments though \cite{exp_zfscaleDTD,exp_zfscaleCHS,exp_zfscaleHTs,exp_zfscaleHLtA}, a description for the ZF-turbulence equilibrium is necessary, which is not provided by contemporary ZF theories.
In the following, the time evolution of the ZFs is investigated and it is shown that the ZF-turbulence equilibrium state is very deterministic. This indicates that the construction of a deterministic RS response functional is feasible. The observations yield a RS response functional that describes the ZF excitation, finite saturation and characteristic radial scale length and permits, for the first time, a reliable charting of ZF-turbulence equilibria.

\begin{figure}[h]
\includegraphics[width=0.48\textwidth]{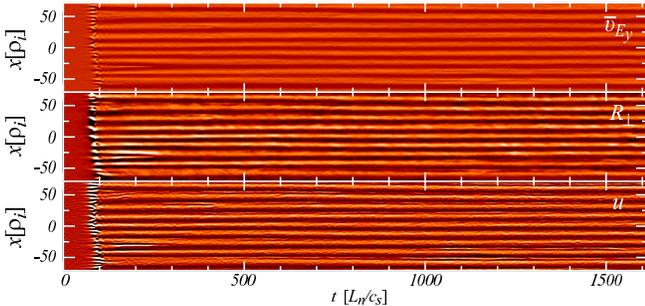}
\vspace{-0.75cm}
\caption{\label{fig:scpattern} Top: self-consistent flow pattern $\vyb=\bar{v}_{E_y}$. Middle:  perp.\ RS $\Rp$. Bottom: shearing rate $u\equiv\dx\vyb$. The red/black color-coding denotes flows in electron/ion diamagnetic drift direction.}
\end{figure}

\paragraph{\label{turbeq}The turbulence equations.---}
The turbulence is described by the electrostatic two-fluid equations assuming adiabatic electrons. The set of equations \cite{code,zfsat} for the potential, temperature and parallel velocity fluctuations, $\phi,T_i$ and $\vpar$ with singly charged ions, the minor and major radii $r$ and $R$, and equal electron and ion background temperatures $T_{e0}$ and $T_{i0}$ is
\begin{align}
  &D_t\left(\phi-\phib \right) -\nap\cdot D_t\nap\left(2\phi-\phib+T_i\right)&\nonumber\\
  &+\dy\phi -\epsilon_n\curv\left(2\phi-\phib+T_i\right)+\epsilon_{\upsilon}\dpar\vpar&=0 \label{eq:phi}\\
  &D_t\left(T_i-2\left(\phi-\phib\right)/3\right)\nonumber\\
  &+\left(\eta_i-2/3\right)\dy\phi-5\epsilon_n\curv T_i/3-2\kappa_i\dpar^2 T_i/3&=0 \label{eq:ti}\\
   &D_t\vpar+\epsilon_{\upsilon}\dpar\left(2\phi+T_i\right)&=0\label{eq:vpar}
\end{align}
The unit for the coordinates  $x$ and $y$ in the radial and poloidal directions is the ion gyro radius, $\rho_i=\sqrt{m_iT_{i0}}/\left(eB\right)$, whereas the parallel coordinate $z\equiv\theta$ ranges from $-\pi\ldots\pi$. The parallel length unit is $L_{\parallel}\equiv qR$ ($q$ is safety factor).
The electron adiabaticity  relation for the density is $n=\phi-\phib$, where the operator $\left<\ldots\right>$ denotes a flux surface average, and $D_t\equiv\partial_t+\left(\hat{z}\times\nap\phi\right)\cdot\nap$. The unit for the fluctuation quantities $e\phi$ and $T_i$ is $T_{i0}\rho_i/L_n$, the velocity unit is $\upsilon_{di}\equiv c_s\rho_i/L_n$ with the ion sound speed $c_s\equiv\sqrt{T_{i0}/m_i}$, and the time unit $t_0\equiv L_n/c_s$. The density and temperature gradient lengths are $L_{n}\equiv dr/d\left(\ln n_0\right)$ and $L_{T_{i}}\equiv dr/d\left(\ln T_{i0}\right)$.
The parallel heat conductivity $\kappa_i$ is chosen to obtain damping rates similar to kinetic phase mixing \cite{zfsat,phase_mix}; $\epsilon_n\equiv 2L_n/R$, $\eta_i\equiv L_n/L_{T_i}$, $\epsilon_{\upsilon}\equiv\epsilon_n/\left(2q\right)$ are dimensionless parameters. For circular high aspect ratio geometry, the curvature operator is $\curv\equiv\cos z ~\dy+\sin z~\dx$ and the parallel derivative is $\dpar\equiv\dz-sx\dy$ for magnetic shear $s$. The mechanisms of ZF generation and saturation described by this system have been addressed in Ref.\ \cite{zfsat} including the ZF shearing properties, decreasing radial transport, and the appearance of a Dimits shift \cite{dimits_shift}.

Stationary ZFs are defined by the wave-number $\Ky=0$, the frequency $\omega=0$ and $\phi=\phib$. Integrating the flux-surface-average of Eq.\ \eqref{eq:phi} over $x$ and subtracting the flux-surface-average of Eq.\ \eqref{eq:vpar} times $-\epsilon_n/\epsilon_v\cos z$  and Eq.\ \eqref{eq:ti} yields an equation for the ZF evolution. Therein the parallel velocity time-derivative is replaced using the return-flow relation $\vpar=-2q\cos z\dx\avg{\phi}$ (required for stationary ZFs to cancel the poloidal flow divergence with the parallel one), which results in
\begin{align}
 \dt\avg{\lrp{1+4q^2\cos(z)^2}\vy}&=-\dx R_t-\xi\label{eq:momentum}
\end{align}
where $\Rt\equiv\Rp-2q\Rpar$ is the total RS and $\xi\equiv 5\epsn\avg{\sin z \dx^2T_i}/3$ a finite larmor radius correction  which is neglected in the following. The perpendicular RS $\Rp=\avg{\vy\left(\vdix+\vx\right)}$ consists of the $\vecE\times\vecB$-velocities $\vy\equiv\dx\phi$, $\vx\equiv-\dy\phi$ and the diamagnetic contribution $\vdix\equiv-\dy\left(\phi-\phib+T_i\right)$. The parallel RS is $\Rpar\equiv\avg{\cos z ~\vpar\vx}$. Alternatively, if one considers $\vy=\avg{\dx\phi+\dx T_i}$ as the ZF velocity then $\Rp=\avg{\vy\vx}$ and $\xi=0$.

\paragraph{\label{numstud}Deterministic flows.---}
A turbulence study with the parameters $\epsilon_n=1.0$, $q=1.5$, $s=1.0$, $\eta_i=2.4$ and a domain of $L_x\times L_y\times L_z=240\ri\times 570\ri\times 2\pi L_{\parallel}$ discretized over a grid $n_x\times n_y\times n_z=512\times 1024\times 32$ shows a nearly stationary poloidal ZF pattern (Fig.\ \ref{fig:scpattern}) with a characteristic scale length that varies only within a small range over time. The ZF-turbulence equilibrium scale length is thereby different from the scale length during the initial ZF excitation phase. States with arbitrary initial flow profiles always decay into the characteristic flow pattern demonstrating the robustness of the radial scale length in the ZF-turbulence equilibrium. Gyro-kinetic turbulence studies using the  GYRO code \cite{gyro} qualitatively reproduce this ZF evolution, which justifies the use of the fluid approach in the following.
For large domains the flow and RS pattern become more deterministic than for smaller ones (e.g.\ $L_y=140\ri$ only) because random fluctuations are averaged out to a greater extent. This determinism permits the following construction of a RS-functional.

\paragraph{Functional construction from observations.---}
Comparison of the RS and ZF patterns in Fig.\ \ref{fig:scpattern} shows that $\Rp$ appears to be proportional to the shearing rate $u\equiv\dx\vyb$.
\begin{figure}[h]
\includegraphics[width=0.48\textwidth]{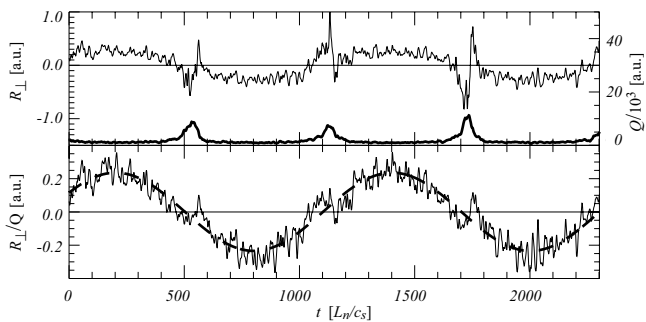}
\vspace{-0.75cm}
\caption{\label{fig:qscale} Time traces at $x=L_x/(4\ri)$. Top: $\Rp$ (thin) and $Q$ (thick). Bottom: $\Rp/Q$ (solid) and $\alzp u$ (dashed).
Comparison shows that $\Rp/Q\approx \alzp u$ with coefficient $\alzp$.}
\end{figure}
A study, with the parameters $\epsilon_n=1.0$, $q=1.4$, $s=0.8$, $\eta_i=3.1$, a domain of $L_x\times L_y\times L_z=122\ri\times 244\ri\times 2\pi L_{\parallel}$ and an artificially maintained flow $\vyb\sim\sin\left(2\pi x\ri/L_x\right)$ that oscillates over time (Fig.\ \ref{fig:qscale}), exhibits variations of at least one order of magnitude in the turbulence intensity $Q\equiv\avyz{\vx T_i}$ which coincide with large deviations of $\Rp$ from $u$.
Rescaling of $\Rp$ by $Q$ using a constant coefficient $\alzp$ evidently restores the proportionality to $u$ .
Comparison of the $Q$ and $u$ profiles reveals further that $Q$ is not just a function of $u$ but follows variations in $u$ with a delay (maxima in $Q$ \emph{after} $u=0$ is reached). This implies that $Q$ should be regarded as a separate degree of freedom for the RS-ZF system.
\begin{figure}[h]
\includegraphics[width=0.48\textwidth]{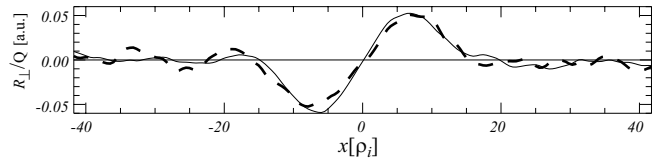}
\vspace{-0.75cm}
\caption{\label{fig:dlnq} Time-average of $\Rp/Q$ (solid) over several flow oscillations compared to the average turbulence intensity gradient $\gamma_{\perp}\dq$ (dashed) with coefficient $\gamma_{\perp}$.}
\end{figure}

To remove all RS contributions caused by $u$, $\Rp/Q$ is time-averaged over several complete flow oscillations.
The residual RS (Fig.\ \ref{fig:dlnq}) is evidently proportional to the time-average of the turbulence intensity gradient $\dq$. 

\begin{figure}[h]
\includegraphics{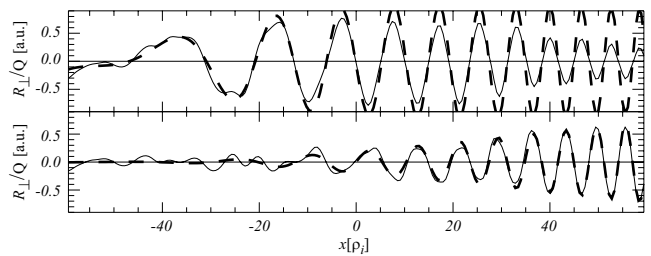}
\vspace{-0.2cm}
\caption{\label{fig:dx2u} Radial profile of an artificial flow study with a radially changing wave-number. Top: Comparison of $\Rp/Q$ (solid) with $u$ (dashed). Bottom: Comparison of $\Rp/Q-\alzp u$ (solid) with $\al_{2,\perp}\dx^2u$ (dashed) with coefficient $\al_{2,\perp}$.}
\end{figure}

To examine the \emph{wave-number dependence} of the stress response a full turbulence study with an artificially enforced flow $\vyb\sim\sin\left(2\pi x\ri(5+15x\ri/L_x)/L_x\right)$ is used (Fig.\ \ref{fig:dx2u}). The parameters are $\epsilon_n=1.0$, $q=1.5$, $s=1.0$, $\eta_i=2.4$ and a domain of $L_x\times L_y\times L_z=140\ri\times 560\ri\times 2\pi L_{\parallel}$.
Apparently the RS decreases with the wave-number $\Kx$ which confirms that the proportionality to $u$ is indeed wave-length dependent.
Since the Eqs.\ \eqref{eq:phi}-\eqref{eq:vpar} are invariant under the transformation $x, y, z, n, \phi, T_i,\vpar, \Rt\rightarrow -x, y, -z, -n,  -\phi, -T_i,\vpar, -\Rt$ the structure of additional terms is restricted.
The symmetry only allows polynomials of even orders in $\Kx$ and a fit shows $\Rp/Q-\alzp u\sim \al_{2,\perp}\dx^2 u$ with a coefficient $\al_{2,\perp}$. 

This symmetry and the previous observations hint that additional terms for the functional are polynomial in $\Kx, u$ and $\dx\ln Q$ as well, thus further simplifying the search for them.

\begin{figure}[h]
\includegraphics{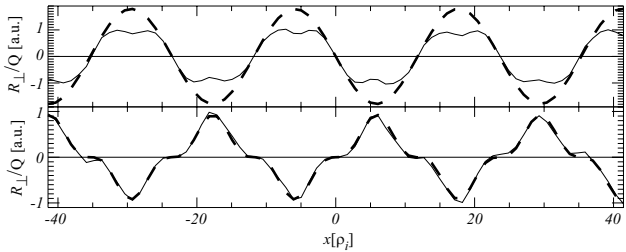}
\vspace{-0.2cm}
\caption{\label{fig:u3} Radial profiles for an artificial flow study with a large amplitude. Top: Comparison of $\Rp/Q$ (solid) with $u$ (dashed). Bottom: Comparison of $\Rp/Q-\alzp u$ (solid) with $\beta_{\perp}u^3$ (dashed) with coefficient $\beta_{\perp}$.}
\end{figure}

The ZF amplitude in self-consistent studies is always finite which indicates a nonlinear saturation term for the functional. To identify the nonlinearity a turbulence study with an artificial flow $\vyb\sim\sin\left(2\pi x\ri/12 L_x \right)$ and a domain of $L_x\times L_y\times L_z=280\ri\times 560\ri\times 2\pi L_{\parallel}$ but otherwise identical parameters to the previous case is used (Fig.\ \ref{fig:u3}). The figure reveals  that $\Rp/Q$ saturates for large shearing rates. The lowest order nonlinear term allowed by the symmetry is $u^3$ and a fit of $u^3$ to $\Rp/Q-\alzp u$ yields the coefficient $\beta_{\perp}$.

Identical terms also describe the parallel stress behavior with coefficients $\{\al_{2i},\beta,\gamma\}_{\parallel}$.
Overall, a functional for the total RS is obtained,
\begin{align}
  \Rt=Q\left(\al_0u-\beta u^3+\al_2\dx^2u -\gamma\dq\right)\, ,
  \label{eq:func_old}
\end{align}
with coefficients $\al_{2i}\equiv\al_{2i,\perp}-2q\al_{2i,\parallel}$, $\beta\equiv\beta_{\perp}-2q\beta_{\parallel}$, $\gamma\equiv\gamma_{\perp}-2q\gamma_{\parallel}$. The coefficients have to be $\al_{2i}, \beta>0$ to fit the observed linear ZF growth and saturation. In addition $\gamma>0$ was found in all turbulence studies.
The functional \eqref{eq:func_old} reproduces the RS  for self-consistent and artificial flow patterns rather accurately including the equilibrium set up phase. However, actual solutions of \eqref{eq:momentum}, using the approximation \eqref{eq:func_old}, a constant $Q$, which is a good approximation in a self-consistent ZF-turbulence equilibrium, and periodic boundary conditions, demonstrate that any initial state successively evolves towards the largest scale length fitting in the system.
This clearly disagrees with the observation of the characteristic and robust ZF scale length (Fig.\ \ref{fig:scpattern}). Apparently one ingredient is missing in the functional \eqref{eq:func_old} to self-consistently describe the ZF evolution.

\paragraph{The missing ingredient.---}
To identify the missing contribution, the ZF behavior induced by the functional \eqref{eq:func_old} is analyzed using a mean-field approximation, $\dx u^3\approx3\meanu\dx u$, where $\left<\ldots \right>_x$ denotes a radial average, which yields a ZF growth rate
\begin{align}
 \Gamma_0\equiv\Kx^2\left(\al_0-3\beta\avx{u^2}-\al_2\Kx^2 \right)\avx{Q}\label{eq:itohgamma}\, .
\end{align}

The region where $\Gamma_0> 0$ is $0<\Kx<\kh$ with $\kh=\sqrt{\al_0-3\beta\avx{u^2}}/\sqrt{\al_2}$. Outside this region flows are damped. $\kh$ tends towards zero for increasing shearing rates $\avx{u^2}$, explaining the eventual ZF decay except for the largest possible scale length.

In contrast, the self-consistent ZF behavior (Fig.\ \ref{fig:scpattern}) requires that small and large $\Kx$ be damped while intermediate $\Kx$ continue to grow until the system saturates at a finite amplitude. To reflect this growth behavior, an additional wave-number dependent term is required to incorporate the necessary additional root to $\Gamma_0=0$ resulting in
\begin{align} 
  \Gamma=\Kx^2\left(\al_0-3\beta\avx{u^2} +\al_2\Kx^2 -\al_4\Kx^4\right)\avx{Q}\,.\label{eq:megamma}
\end{align}
This formula confines the region of growth to a band $\kl<\Kx<\kh$, where $0<\kl<\kh$ are the roots of $\Gamma=0$, for sufficiently high shearing rates and $\al_4>0$. Since $\kl$ increases and $\kh$ decreases with the shearing rate, the system saturates at $\kl=\kh=\sqrt{\al_2/2\al_4}$. The corresponding ZF evolution equation is
\begin{align}
 \dt u=-\dx^2&\left[Q\left(\al_0 u-\beta u^3 -\al_2\dx^2u\right.\right. \label{eq:mefunc}\\
       &\left.\left.-\gamma\dq -\al_4\dx^4u\right)\right]\nonumber\, .
\end{align}
Numerical solutions of Eq.\ \eqref{eq:mefunc} always yield a stable state with a finite amplitude and scale length as required by the phenomenology of the RS-ZF-system, corroborating the mean field theory.
\paragraph{Measurement of the $\Kx^4$-term.---}
Unfortunately, despite the very deterministic RS pattern, a least-squares fit of \eqref{eq:mefunc} to the turbulence runs still proved to have unacceptably large errors for the coefficient $\al_4$. To verify the restriction of the ZF-drive to a wave-number band required by mean-field theory, a scenario with a fixed background shearing-rate $u_p=const$ (providing the mean-field component $\avx{u^2}\approx\avx{u_p^2}$) and a small perturbatory shearing-rate $u_s\sim\sin\left(2\pi x\ri\Kx\right)$, $\avx{u_s^2}^{\scriptstyle 1/2}=0.1\avx{u_p^2}^{\scriptstyle 1/2}$, with varying $\Kx$ is studied. In addition an optimal filtering technique is employed to measure the stress response to $u_s$.

Writing the time-average of $\Rt/Q$ as
\begin{align}
  \overline{\lrp{\frac{\Rt}{Q}}}\lrp{x}=\sum\limits_i\tilde{\al}_iP_i\lrp{x}+\sum\limits_j\beta_jN_j\lrp{x}+n\lrp{x}
\end{align}
where the $P_i$ form a set of Ansatz functionals $\lrp{P_i\lrp{x}}=\lrp{u_s,u_p,\dx\ln Q}$ whereas the $N_j\lrp{x}$ constitute a set of possible error terms $\lrp{N_j\lrp{x}}=\lrp{x,x^2}$ and $n\lrp{x}$ represents random noise with coefficients $\tilde{\al}_i,\beta_j$. We estimate $\tal_1$ with
\begin{align}
 \left.\tal_1\right|_{\mathnormal{est}}&\equiv\sum\limits_{i,j}\overline{\lrp{\frac{\Rt}{Q}}}\lrp{x_i}\lrp{C^{-1}}_{ij}P_1\lrp{x_j}/S\\
  S&\equiv\sum_{i,j}P_1\lrp{x_i}\lrp{C^{-1}}_{ij}P_1\lrp{x_j}\,
\end{align}
using a minimal variance estimator.
The required covariance matrix is initially estimated by
\begin{align}
 C_{ij}=v_{bg}+\sum\limits_l v_{l,P}\hat{P}_{l,i}\hat{P}_{l,j}+\sum\limits_l v_{l,N}\hat{N}_{l,i}\hat{N}_{l,j}
\end{align}
where $\hat{P}_{l,i}\equiv P_l\lrp{x_i}/\mathnormal{max}\lrp{P_l}$, $\hat{N}_{l,i}\equiv N_l\lrp{x_i}/\mathnormal{max}\lrp{N_l}$ with $v_{bg}, v_{l,P}, v_{l,N}$ are ``a priori'' estimates for the variances obtained from observations in a large ensemble of turbulence studies. The covariance is iteratively refined using the variances computed from coefficient estimates at different times.

\begin{figure}[h]
\includegraphics{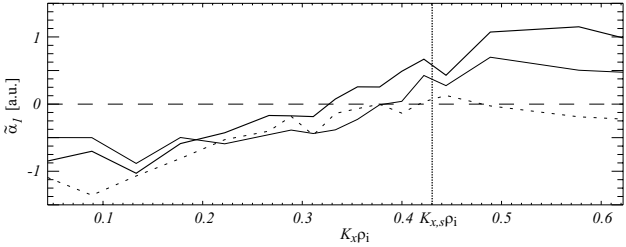}
\vspace{-0.2cm}
\caption{\label{fig:gscalemeasure} $\Kx$-dependence of $\tal_1$ in an artificial-flow study with $u=u_p+u_s$ for different primary shearing rates $u_p$ equal to $0.1/0.12/0.15\, [a.u.]$ (thick/thin/dashed). Self-consistent ZF wave-number is $K_{x,s}$.}
\end{figure}

Figure \ref{fig:gscalemeasure} shows measurements of $\tal_1$ for several shearing rates $u_p$.
Evidently shearing-rates $u_s$ with small $\Kx$ are damped and the region where $\tal_1>0$ has a lower boundary $\kl$ which increases for larger $u_p$ until saturation, a behavior like the one described by $\Gamma/\Kx^2$, thus validating formula \eqref{eq:mefunc}.

\paragraph{\label{con}Discussion---}
The response functional for the total RS is
\begin{align}
 \Rt&=Q\left(\al_0 u+\beta u^3+\al_2\dx^2u\right.\nonumber\\
&\left.+\gamma\dq +\al_4\dx^4u\right)\label{eq:Rfunc},
\end{align}
with coefficients $\al_{2i}\equiv\al_{2i,\perp}-2q\al_{2i,\parallel}$, $\beta\equiv\beta_{\perp}-2q\beta_{\parallel}$, $\gamma\equiv\gamma_{\perp}-2q\gamma_{\parallel}$, $\al_0>0$, $\al_{2i>0},\beta,\gamma<0$.
The functional reproduces the features observed in self-consistent ZF patterns (Fig.\ \ref{fig:scpattern}) and describes the excitation, finite saturation and robust characteristic radial scale length of the ZFs.

To gain analytical insight on the coefficients it is instructive to use wave-kinetics,  in a drift-wave model system 
(although strictly speaking there is no radial scale separation, a prerequisite for wave-kinetic theory \cite{wk_derivation}, between the ITG-turbulence driving the ZF and the ZFs themselves). This has been carried out for Eq.\ \eqref{eq:func_old} (without $\dx\ln Q$) in \cite{itoh_hallatschek_zfwavek,todaItoh_zfscale,smolyakov_wk,Itoh2005}. The wave-kinetic equation is
\begin{equation}
 \dt\Nk+\nak\omega\cdot\nax\Nk-\nax\omega\cdot\nak\Nk=C\left (\Nk\right)~,
 \label{eq:waveaction}
\end{equation}
where the adiabatic wave-action is $\Nk\propto \nfac\left|\phi\right|^2_{\boldsymbol{k}}$ with a symmetric $\boldsymbol{k}$-dependent coefficient $\nfac$. For drift-waves (DW) $\nfac=(1+\kp^2)^2$ \cite{itoh_hallatschek_zfwavek,smolyakov_wk,Dyachenko_wkintro,LebedevDiamond_wkintro}. 
The dispersion $\omega=\odw+\ky\vy$ consists of the local frequency $\odw$ [$\odw=\ky/\left(1+\kp^2\right)$ for DW] and the Doppler shift $\ky\vy$. The ``collisional''-term $C\left (\Nk\right)=\cop\left(\Nkz -\Nk\right)$ describes tendency of $\Nk$ to evolve into a turbulence equilibrium state $\Nkz$. $\Rp$ is then given by 
\begin{align}
   \Rp=\left<\vx\vy\right>={\textstyle\int}\dtk\pik\avyz{\Nk}
 \label{eq:Rpwk}
\end{align}
where $\pik\equiv \kx\ky/\nfac$. 

An expansion of $\Nk$ with respect the various orders of wave interactions $u^m,\Kx^n u\equiv\dx^n u$ ($m,n\in\mathbb{N}$) and $\dx\ln Q$ \cite{itoh_hallatschek_zfwavek,smolyakov_wk} and terms of even higher order solves Eq.\ \eqref{eq:waveaction} and one can calculate an estimate for $\Rp$ assuming $\dt\ll\cop$:

\begin{align}
 \Rp/Q=\sum\limits_{i=0}^n\underbrace{ a_{2i}\dx^{2i}u}_{u^1,\Kx^i}+\underbrace{b u^3}_{u^3,\Kx^0}+c\dx\ln Q ~.
 \label{eq:rpest}
\end{align}
For $\Nk\propto Q$ and $\dx\Nkz\approx(\Nkz/Q)\dx Q$ the coefficients
$a_i,c_0,d_1$ are independent of $Q$ and given by 
\begin{align}
    a_{2i}&=\avk{\dkx\left[\pik\ky\vgx^{2i}\right]}/\cop^{2i+1}\label{eq:ai}\\
 b&=\avk{\ky^3\dkx^3\pik}/\cop^3\label{eq:c0}\\
   c&=\avk{\pik\vgx}/\cop\label{eq:d1}
\end{align}
with $\avk{A}\equiv\int\dtk A\avyz{\Nkz}/Q$. 
\begin{table}[h]
% in PRL units
 \begin{tabular}{l|c|c|c|c|c}
  &$\al_{0,*}|a_0$&$\beta|b$&$\al_{2,*}|a_2$&$\al_{4,*}|a_4$&$\gamma|c$\\
  \hline
  $\Rp$ meas.&$2.1$&$-3.1$&$1.3$&$5.1$&$-2.1$\\
  \hline
  $\Rp$ w.-k.&$2.5$&$-3.5$&$0.2$&$2\cdot 10^{-3}$&$-0.82$\\
  \hline
  $\Rpar$ meas.&$0.56$&$-1.0$&$0.84$&$2.7$&$-0.13$
 \end{tabular}
 \caption{Measured and wave-kinetically derived coefficients.}
  \label{tab:rperp}
\end{table}
The numerical values for the coefficients for $\Rp$ are shown in Table \ref{tab:rperp}. The signs of the wave-kinetic coefficients agree with the measurements for $\Rp$. However, using only the coefficients $\al_{0-4,\perp}$ and $\al_{0,\parallel}$ as discussed in \cite{itoh_hallatschek_zfwavek,todaItoh_zfscale,smolyakov_wk,Itoh2005} results in a total stress functional that always leads to a final ZF pattern with the largest wave-length admissible by the boundary conditions  (see discussion of Eq.\ \eqref{eq:func_old}). The additional term represents a limitation of the flow wave-length, which becomes effective at flow amplitudes approaching the self-consistent ones but is \emph{absent} for small amplitudes. It is therefore essential to take the higher order contributions by $\Rpar$ and by $\dx^4u$ to $\Rt$ into account, a fact that has been largely neglected in all contemporary wave-kinetic ZF theories.

\paragraph{\label{outlook}Outlook---}
The discussed measurement technique can, in principle, be applied to obtain the coefficient dependencies on the plasma parameters or to investigate the initial conditions of turbulence triggered transport barriers. Further studies of the stress response may reveal additional higher-order wave-number-dependent or nonlinear terms that describe metastable ZF states. This would allow a reliable charting of ZF-turbulence equilibria opposite to the standard procedure of turbulence parameter scans where the metastability might not be identified.

\bibliography{PRL}% Produces the bibliography via BibTeX.

\providecommand{\noopsort}[1]{}\providecommand{\singleletter}[1]{#1}%
\begin{thebibliography}{27}%
\makeatletter
\providecommand \@ifxundefined [1]{%
 \@ifx{#1\undefined}
}%
\providecommand \@ifnum [1]{%
 \ifnum #1\expandafter \@firstoftwo
 \else \expandafter \@secondoftwo
 \fi
}%
\providecommand \@ifx [1]{%
 \ifx #1\expandafter \@firstoftwo
 \else \expandafter \@secondoftwo
 \fi
}%
\providecommand \natexlab [1]{#1}%
\providecommand \enquote  [1]{``#1''}%
\providecommand \bibnamefont  [1]{#1}%
\providecommand \bibfnamefont [1]{#1}%
\providecommand \citenamefont [1]{#1}%
\providecommand \href@noop [0]{\@secondoftwo}%
\providecommand \href [0]{\begingroup \@sanitize@url \@href}%
\providecommand \@href[1]{\@@startlink{#1}\@@href}%
\providecommand \@@href[1]{\endgroup#1\@@endlink}%
\providecommand \@sanitize@url [0]{\catcode `\\12\catcode `\$12\catcode
  `\&12\catcode `\#12\catcode `\^12\catcode `\_12\catcode `\%12\relax}%
\providecommand \@@startlink[1]{}%
\providecommand \@@endlink[0]{}%
\providecommand \url  [0]{\begingroup\@sanitize@url \@url }%
\providecommand \@url [1]{\endgroup\@href {#1}{\urlprefix }}%
\providecommand \urlprefix  [0]{URL }%
\providecommand \Eprint [0]{\href }%
\providecommand \doibase [0]{http://dx.doi.org/}%
\providecommand \selectlanguage [0]{\@gobble}%
\providecommand \bibinfo  [0]{\@secondoftwo}%
\providecommand \bibfield  [0]{\@secondoftwo}%
\providecommand \translation [1]{[#1]}%
\providecommand \BibitemOpen [0]{}%
\providecommand \bibitemStop [0]{}%
\providecommand \bibitemNoStop [0]{.\EOS\space}%
\providecommand \EOS [0]{\spacefactor3000\relax}%
\providecommand \BibitemShut  [1]{\csname bibitem#1\endcsname}%
\let\auto@bib@innerbib\@empty
%</preamble>
\bibitem [{\citenamefont {Lin}\ \emph {et~al.}(2000)\citenamefont {Lin},
  \citenamefont {Hahm}, \citenamefont {Lee}, \citenamefont {Tang},\ and\
  \citenamefont {White}}]{linhahm_zfshear}%
  \BibitemOpen
  \bibfield  {author} {\bibinfo {author} {\bibfnamefont {Z.}~\bibnamefont
  {Lin}}, \bibinfo {author} {\bibfnamefont {T.}~\bibnamefont {Hahm}}, \bibinfo
  {author} {\bibfnamefont {W.}~\bibnamefont {Lee}}, \bibinfo {author}
  {\bibfnamefont {W.}~\bibnamefont {Tang}}, \ and\ \bibinfo {author}
  {\bibfnamefont {R.}~\bibnamefont {White}},\ }\href@noop {} {\bibfield
  {journal} {\bibinfo  {journal} {Phys. Plasmas}\ }\textbf {\bibinfo {volume}
  {7}},\ \bibinfo {pages} {1857} (\bibinfo {year} {2000})}\BibitemShut
  {NoStop}%
\bibitem [{\citenamefont {Rosenbluth}\ and\ \citenamefont
  {Hinton}(1998)}]{rosenbHinton_zfshear}%
  \BibitemOpen
  \bibfield  {author} {\bibinfo {author} {\bibfnamefont {M.}~\bibnamefont
  {Rosenbluth}}\ and\ \bibinfo {author} {\bibfnamefont {F.}~\bibnamefont
  {Hinton}},\ }\href@noop {} {\bibfield  {journal} {\bibinfo  {journal} {Phys.
  Rev. Lett.}\ }\textbf {\bibinfo {volume} {80}},\ \bibinfo {pages} {724}
  (\bibinfo {year} {1998})}\BibitemShut {NoStop}%
\bibitem [{\citenamefont {Hammett}\ \emph {et~al.}(1993)\citenamefont
  {Hammett}, \citenamefont {Beer}, \citenamefont {Dorland}, \citenamefont
  {Cowley},\ and\ \citenamefont {Smith}}]{hammett_zfshear}%
  \BibitemOpen
  \bibfield  {author} {\bibinfo {author} {\bibfnamefont {G.}~\bibnamefont
  {Hammett}}, \bibinfo {author} {\bibfnamefont {M.~A.}\ \bibnamefont {Beer}},
  \bibinfo {author} {\bibfnamefont {W.}~\bibnamefont {Dorland}}, \bibinfo
  {author} {\bibfnamefont {S.}~\bibnamefont {Cowley}}, \ and\ \bibinfo {author}
  {\bibfnamefont {S.}~\bibnamefont {Smith}},\ }\href@noop {} {\bibfield
  {journal} {\bibinfo  {journal} {PPCF}\ }\textbf {\bibinfo {volume} {35}},\
  \bibinfo {pages} {973} (\bibinfo {year} {1993})}\BibitemShut {NoStop}%
\bibitem [{\citenamefont {Okuda}\ \emph {et~al.}(1980)\citenamefont {Okuda},
  \citenamefont {Sato}, \citenamefont {Hasegawa},\ and\ \citenamefont
  {Pellat}}]{hasegawa_zfgen}%
  \BibitemOpen
  \bibfield  {author} {\bibinfo {author} {\bibfnamefont {H.}~\bibnamefont
  {Okuda}}, \bibinfo {author} {\bibfnamefont {T.}~\bibnamefont {Sato}},
  \bibinfo {author} {\bibfnamefont {A.}~\bibnamefont {Hasegawa}}, \ and\
  \bibinfo {author} {\bibfnamefont {R.}~\bibnamefont {Pellat}},\ }\href@noop {}
  {\bibfield  {journal} {\bibinfo  {journal} {Phys. Fluids}\ }\textbf {\bibinfo
  {volume} {23}},\ \bibinfo {pages} {10} (\bibinfo {year} {1980})}\BibitemShut
  {NoStop}%
\bibitem [{\citenamefont {Diamond}\ and\ \citenamefont
  {Kim}(1991)}]{diamondKim_zfgen}%
  \BibitemOpen
  \bibfield  {author} {\bibinfo {author} {\bibfnamefont {P.}~\bibnamefont
  {Diamond}}\ and\ \bibinfo {author} {\bibfnamefont {Y.}~\bibnamefont {Kim}},\
  }\href@noop {} {\bibfield  {journal} {\bibinfo  {journal} {Phys. Fluids B}\
  }\textbf {\bibinfo {volume} {3}},\ \bibinfo {pages} {1626} (\bibinfo {year}
  {1991})}\BibitemShut {NoStop}%
\bibitem [{\citenamefont {Diamond}\ \emph {et~al.}(2007)\citenamefont
  {Diamond}, \citenamefont {Rosenbluth}, \citenamefont {Hinton}, \citenamefont
  {Malkov}, \citenamefont {Fleischer},\ and\ \citenamefont
  {Smolyakov}}]{diamondRosenbluth_IAEA_zfgen}%
  \BibitemOpen
  \bibinfo {editor} {\bibfnamefont {P.}~\bibnamefont {Diamond}}, \bibinfo
  {editor} {\bibfnamefont {M.}~\bibnamefont {Rosenbluth}}, \bibinfo {editor}
  {\bibfnamefont {F.}~\bibnamefont {Hinton}}, \bibinfo {editor} {\bibfnamefont
  {M.}~\bibnamefont {Malkov}}, \bibinfo {editor} {\bibfnamefont
  {J.}~\bibnamefont {Fleischer}}, \ and\ \bibinfo {editor} {\bibfnamefont
  {A.}~\bibnamefont {Smolyakov}},\ eds.,\ \href@noop {} {\emph {\bibinfo
  {title} {17th IAEA Proceedings}}}\ (\bibinfo {year} {2007})\BibitemShut
  {NoStop}%
\bibitem [{\citenamefont {Itoh}\ \emph {et~al.}(2006)\citenamefont {Itoh},
  \citenamefont {Itoh}, \citenamefont {Diamond}, \citenamefont {Hahm},
  \citenamefont {Fujisawa}, \citenamefont {Tynan}, \citenamefont {Yagi},\ and\
  \citenamefont {Nagashima}}]{itoh_zfphysics}%
  \BibitemOpen
  \bibfield  {author} {\bibinfo {author} {\bibfnamefont {K.}~\bibnamefont
  {Itoh}}, \bibinfo {author} {\bibfnamefont {S.-I.}\ \bibnamefont {Itoh}},
  \bibinfo {author} {\bibfnamefont {P.}~\bibnamefont {Diamond}}, \bibinfo
  {author} {\bibfnamefont {T.}~\bibnamefont {Hahm}}, \bibinfo {author}
  {\bibfnamefont {A.}~\bibnamefont {Fujisawa}}, \bibinfo {author}
  {\bibfnamefont {G.}~\bibnamefont {Tynan}}, \bibinfo {author} {\bibfnamefont
  {M.}~\bibnamefont {Yagi}}, \ and\ \bibinfo {author} {\bibfnamefont
  {Y.}~\bibnamefont {Nagashima}},\ }\href@noop {} {\bibfield  {journal}
  {\bibinfo  {journal} {Phys. Plasmas}\ }\textbf {\bibinfo {volume} {13}}
  (\bibinfo {year} {2006})}\BibitemShut {NoStop}%
\bibitem [{\citenamefont {Itoh}\ \emph {et~al.}(2004)\citenamefont {Itoh},
  \citenamefont {Hallatschek}, \citenamefont {Toda}, \citenamefont {Sanuki},\
  and\ \citenamefont {Itoh}}]{itoh_hallatschek_zfwavek}%
  \BibitemOpen
  \bibfield  {author} {\bibinfo {author} {\bibfnamefont {K.}~\bibnamefont
  {Itoh}}, \bibinfo {author} {\bibfnamefont {K.}~\bibnamefont {Hallatschek}},
  \bibinfo {author} {\bibfnamefont {S.}~\bibnamefont {Toda}}, \bibinfo {author}
  {\bibfnamefont {H.}~\bibnamefont {Sanuki}}, \ and\ \bibinfo {author}
  {\bibfnamefont {S.-I.}\ \bibnamefont {Itoh}},\ }\href@noop {} {\bibfield
  {journal} {\bibinfo  {journal} {J. Phys. Soc. Japan}\ }\textbf {\bibinfo
  {volume} {73}},\ \bibinfo {pages} {2921} (\bibinfo {year}
  {2004})}\BibitemShut {NoStop}%
\bibitem [{\citenamefont {Hallatschek}\ and\ \citenamefont
  {Diamond}(2003)}]{diamondHallatschek_instabdrift}%
  \BibitemOpen
  \bibfield  {author} {\bibinfo {author} {\bibfnamefont {K.}~\bibnamefont
  {Hallatschek}}\ and\ \bibinfo {author} {\bibfnamefont {P.}~\bibnamefont
  {Diamond}},\ }\href@noop {} {\bibfield  {journal} {\bibinfo  {journal}
  {NJPH}\ }\textbf {\bibinfo {volume} {5}},\ \bibinfo {pages} {29.1} (\bibinfo
  {year} {2003})}\BibitemShut {NoStop}%
\bibitem [{\citenamefont {Itoh}\ \emph {et~al.}(2005)\citenamefont {Itoh},
  \citenamefont {Hallatschek}, \citenamefont {Itoh}, \citenamefont {Diamond},\
  and\ \citenamefont {Toda}}]{Itoh2005}%
  \BibitemOpen
  \bibfield  {author} {\bibinfo {author} {\bibfnamefont {K.}~\bibnamefont
  {Itoh}}, \bibinfo {author} {\bibfnamefont {K.}~\bibnamefont {Hallatschek}},
  \bibinfo {author} {\bibfnamefont {S.-I.}\ \bibnamefont {Itoh}}, \bibinfo
  {author} {\bibfnamefont {P.~H.}\ \bibnamefont {Diamond}}, \ and\ \bibinfo
  {author} {\bibfnamefont {S.}~\bibnamefont {Toda}},\ }\href {\doibase
  10.1063/1.1922788} {\bibfield  {journal} {\bibinfo  {journal} {Physics of
  Plasmas}\ }\textbf {\bibinfo {volume} {12}},\ \bibinfo {pages} {062303}
  (\bibinfo {year} {2005})}\BibitemShut {NoStop}%
\bibitem [{\citenamefont {Hallatschek}\ and\ \citenamefont
  {Biskamp}(2001)}]{gamprl}%
  \BibitemOpen
  \bibfield  {author} {\bibinfo {author} {\bibfnamefont {K.}~\bibnamefont
  {Hallatschek}}\ and\ \bibinfo {author} {\bibfnamefont {D.}~\bibnamefont
  {Biskamp}},\ }\href@noop {} {\bibfield  {journal} {\bibinfo  {journal} {Phys.
  Rev. Let.}\ }\textbf {\bibinfo {volume} {86}},\ \bibinfo {pages} {7}
  (\bibinfo {year} {2001})}\BibitemShut {NoStop}%
\bibitem [{\citenamefont {Rogers}\ \emph {et~al.}(2000)\citenamefont {Rogers},
  \citenamefont {Dorland},\ and\ \citenamefont
  {M.Kotschenreuther}}]{dorland_zfinstab}%
  \BibitemOpen
  \bibfield  {author} {\bibinfo {author} {\bibfnamefont {B.}~\bibnamefont
  {Rogers}}, \bibinfo {author} {\bibfnamefont {W.}~\bibnamefont {Dorland}}, \
  and\ \bibinfo {author} {\bibnamefont {M.Kotschenreuther}},\ }\href@noop {}
  {\bibfield  {journal} {\bibinfo  {journal} {Phys. Rev. Let.}\ }\textbf
  {\bibinfo {volume} {85}},\ \bibinfo {pages} {25} (\bibinfo {year}
  {2000})}\BibitemShut {NoStop}%
\bibitem [{\citenamefont {Lin}\ \emph {et~al.}(1999)\citenamefont {Lin},
  \citenamefont {Hahm}, \citenamefont {Lee}, \citenamefont {Tang},\ and\
  \citenamefont {Diamond}}]{linhahm_zfinstab}%
  \BibitemOpen
  \bibfield  {author} {\bibinfo {author} {\bibfnamefont {Z.}~\bibnamefont
  {Lin}}, \bibinfo {author} {\bibfnamefont {T.}~\bibnamefont {Hahm}}, \bibinfo
  {author} {\bibfnamefont {W.}~\bibnamefont {Lee}}, \bibinfo {author}
  {\bibfnamefont {W.}~\bibnamefont {Tang}}, \ and\ \bibinfo {author}
  {\bibfnamefont {P.}~\bibnamefont {Diamond}},\ }\href@noop {} {\bibfield
  {journal} {\bibinfo  {journal} {Phys. Rev. Let.}\ }\textbf {\bibinfo {volume}
  {83}},\ \bibinfo {pages} {18} (\bibinfo {year} {1999})}\BibitemShut {NoStop}%
\bibitem [{\citenamefont {Gupta}\ \emph {et~al.}(2006)\citenamefont {Gupta},
  \citenamefont {Fonck}, \citenamefont {McKee}, \citenamefont {Schlossberg},\
  and\ \citenamefont {Shafer}}]{exp_zfscaleDTD}%
  \BibitemOpen
  \bibfield  {author} {\bibinfo {author} {\bibfnamefont {D.}~\bibnamefont
  {Gupta}}, \bibinfo {author} {\bibfnamefont {R.}~\bibnamefont {Fonck}},
  \bibinfo {author} {\bibfnamefont {G.}~\bibnamefont {McKee}}, \bibinfo
  {author} {\bibfnamefont {D.}~\bibnamefont {Schlossberg}}, \ and\ \bibinfo
  {author} {\bibfnamefont {M.}~\bibnamefont {Shafer}},\ }\href@noop {}
  {\bibfield  {journal} {\bibinfo  {journal} {Phys. Rev. Lett.}\ }\textbf
  {\bibinfo {volume} {97}},\ \bibinfo {pages} {125002} (\bibinfo {year}
  {2006})}\BibitemShut {NoStop}%
\bibitem [{\citenamefont {Fujisawa}\ \emph {et~al.}(2004)\citenamefont
  {Fujisawa}, \citenamefont {Itoh}, \citenamefont {Iguchi}, \citenamefont
  {Matsuoka}, \citenamefont {Okamura}, \citenamefont {Shimizu}, \citenamefont
  {Minami}, \citenamefont {Yoshimura}, \citenamefont {Nagaoka}, \citenamefont
  {Takahashi}, \citenamefont {Kojima}, \citenamefont {Nakano}, \citenamefont
  {Ohsima}, \citenamefont {Nishimura}, \citenamefont {Isobe}, \citenamefont
  {Suzuki}, \citenamefont {Akiyama}, \citenamefont {Ida}, \citenamefont {Toi},
  \citenamefont {Itoh},\ and\ \citenamefont {Diamond}}]{exp_zfscaleCHS}%
  \BibitemOpen
  \bibfield  {author} {\bibinfo {author} {\bibfnamefont {A.}~\bibnamefont
  {Fujisawa}}, \bibinfo {author} {\bibfnamefont {K.}~\bibnamefont {Itoh}},
  \bibinfo {author} {\bibfnamefont {H.}~\bibnamefont {Iguchi}}, \bibinfo
  {author} {\bibfnamefont {K.}~\bibnamefont {Matsuoka}}, \bibinfo {author}
  {\bibfnamefont {S.}~\bibnamefont {Okamura}}, \bibinfo {author} {\bibfnamefont
  {A.}~\bibnamefont {Shimizu}}, \bibinfo {author} {\bibfnamefont
  {T.}~\bibnamefont {Minami}}, \bibinfo {author} {\bibfnamefont
  {Y.}~\bibnamefont {Yoshimura}}, \bibinfo {author} {\bibfnamefont
  {K.}~\bibnamefont {Nagaoka}}, \bibinfo {author} {\bibfnamefont
  {C.}~\bibnamefont {Takahashi}}, \bibinfo {author} {\bibfnamefont
  {M.}~\bibnamefont {Kojima}}, \bibinfo {author} {\bibfnamefont
  {H.}~\bibnamefont {Nakano}}, \bibinfo {author} {\bibfnamefont
  {S.}~\bibnamefont {Ohsima}}, \bibinfo {author} {\bibfnamefont
  {S.}~\bibnamefont {Nishimura}}, \bibinfo {author} {\bibfnamefont
  {M.}~\bibnamefont {Isobe}}, \bibinfo {author} {\bibfnamefont
  {C.}~\bibnamefont {Suzuki}}, \bibinfo {author} {\bibfnamefont
  {T.}~\bibnamefont {Akiyama}}, \bibinfo {author} {\bibfnamefont
  {K.}~\bibnamefont {Ida}}, \bibinfo {author} {\bibfnamefont {K.}~\bibnamefont
  {Toi}}, \bibinfo {author} {\bibfnamefont {S.-I.}\ \bibnamefont {Itoh}}, \
  and\ \bibinfo {author} {\bibfnamefont {P.}~\bibnamefont {Diamond}},\
  }\href@noop {} {\bibfield  {journal} {\bibinfo  {journal} {Phys. Rev. Lett.}\
  }\textbf {\bibinfo {volume} {93}},\ \bibinfo {pages} {165002} (\bibinfo
  {year} {2004})}\BibitemShut {NoStop}%
\bibitem [{\citenamefont {Xu}\ \emph {et~al.}(2003)\citenamefont {Xu},
  \citenamefont {Wan}, \citenamefont {Song},\ and\ \citenamefont
  {Li}}]{exp_zfscaleHTs}%
  \BibitemOpen
  \bibfield  {author} {\bibinfo {author} {\bibfnamefont {G.}~\bibnamefont
  {Xu}}, \bibinfo {author} {\bibfnamefont {B.}~\bibnamefont {Wan}}, \bibinfo
  {author} {\bibfnamefont {M.}~\bibnamefont {Song}}, \ and\ \bibinfo {author}
  {\bibfnamefont {J.}~\bibnamefont {Li}},\ }\href@noop {} {\bibfield  {journal}
  {\bibinfo  {journal} {Phys. Rev. Lett.}\ }\textbf {\bibinfo {volume} {91}},\
  \bibinfo {pages} {125001} (\bibinfo {year} {2003})}\BibitemShut {NoStop}%
\bibitem [{\citenamefont {Liu}\ \emph {et~al.}(2009)\citenamefont {Liu},
  \citenamefont {Lan}, \citenamefont {Yu}, \citenamefont {Zhao}, \citenamefont
  {Yan}, \citenamefont {Hong}, \citenamefont {Dong}, \citenamefont {Zhao},
  \citenamefont {Qian}, \citenamefont {Cheng}, \citenamefont {Duan},\ and\
  \citenamefont {Liu}}]{exp_zfscaleHLtA}%
  \BibitemOpen
  \bibfield  {author} {\bibinfo {author} {\bibfnamefont {A.~D.}\ \bibnamefont
  {Liu}}, \bibinfo {author} {\bibfnamefont {T.}~\bibnamefont {Lan}}, \bibinfo
  {author} {\bibfnamefont {C.~X.}\ \bibnamefont {Yu}}, \bibinfo {author}
  {\bibfnamefont {H.~L.}\ \bibnamefont {Zhao}}, \bibinfo {author}
  {\bibfnamefont {L.~W.}\ \bibnamefont {Yan}}, \bibinfo {author} {\bibfnamefont
  {W.~Y.}\ \bibnamefont {Hong}}, \bibinfo {author} {\bibfnamefont {J.~Q.}\
  \bibnamefont {Dong}}, \bibinfo {author} {\bibfnamefont {K.~J.}\ \bibnamefont
  {Zhao}}, \bibinfo {author} {\bibfnamefont {J.}~\bibnamefont {Qian}}, \bibinfo
  {author} {\bibfnamefont {J.}~\bibnamefont {Cheng}}, \bibinfo {author}
  {\bibfnamefont {X.~R.}\ \bibnamefont {Duan}}, \ and\ \bibinfo {author}
  {\bibfnamefont {Y.}~\bibnamefont {Liu}},\ }\href@noop {} {\bibfield
  {journal} {\bibinfo  {journal} {Phys. Rev. Lett.}\ }\textbf {\bibinfo
  {volume} {103}},\ \bibinfo {pages} {095002} (\bibinfo {year}
  {2009})}\BibitemShut {NoStop}%
\bibitem [{\citenamefont {Hallatschek}\ and\ \citenamefont
  {Zeiler}(2000)}]{code}%
  \BibitemOpen
  \bibfield  {author} {\bibinfo {author} {\bibfnamefont {K.}~\bibnamefont
  {Hallatschek}}\ and\ \bibinfo {author} {\bibfnamefont {A.}~\bibnamefont
  {Zeiler}},\ }\href@noop {} {\bibfield  {journal} {\bibinfo  {journal} {Phys.
  Plasmas}\ }\textbf {\bibinfo {volume} {7}},\ \bibinfo {pages} {2554}
  (\bibinfo {year} {2000})}\BibitemShut {NoStop}%
\bibitem [{\citenamefont {Hallatschek}(2004)}]{zfsat}%
  \BibitemOpen
  \bibfield  {author} {\bibinfo {author} {\bibfnamefont {K.}~\bibnamefont
  {Hallatschek}},\ }\href@noop {} {\bibfield  {journal} {\bibinfo  {journal}
  {Phys. Rev. Let.}\ }\textbf {\bibinfo {volume} {93}},\ \bibinfo {pages} {6}
  (\bibinfo {year} {2004})}\BibitemShut {NoStop}%
\bibitem [{\citenamefont {Hammett}\ and\ \citenamefont
  {Perkins}(1990)}]{phase_mix}%
  \BibitemOpen
  \bibfield  {author} {\bibinfo {author} {\bibfnamefont {G.~W.}\ \bibnamefont
  {Hammett}}\ and\ \bibinfo {author} {\bibfnamefont {F.~W.}\ \bibnamefont
  {Perkins}},\ }\href@noop {} {\bibfield  {journal} {\bibinfo  {journal} {Phys.
  Rev. Lett.}\ }\textbf {\bibinfo {volume} {64}},\ \bibinfo {pages} {25}
  (\bibinfo {year} {1990})}\BibitemShut {NoStop}%
\bibitem [{\citenamefont {Dimits}\ \emph {et~al.}(2000)\citenamefont {Dimits},
  \citenamefont {Bateman}, \citenamefont {Beer}, \citenamefont {Cohen},
  \citenamefont {Dorland}, \citenamefont {Hammett}, \citenamefont {Kim},
  \citenamefont {Kinsey}, \citenamefont {Kotschenreuther}, \citenamefont
  {Kritz}, \citenamefont {Lao}, \citenamefont {Mandrekas}, \citenamefont
  {Nevins}, \citenamefont {Parker}, \citenamefont {Redd}, \citenamefont
  {Shumaker}, \citenamefont {R.Sydora},\ and\ \citenamefont
  {Weiland}}]{dimits_shift}%
  \BibitemOpen
  \bibfield  {author} {\bibinfo {author} {\bibfnamefont {A.}~\bibnamefont
  {Dimits}}, \bibinfo {author} {\bibfnamefont {G.}~\bibnamefont {Bateman}},
  \bibinfo {author} {\bibfnamefont {M.}~\bibnamefont {Beer}}, \bibinfo {author}
  {\bibfnamefont {B.}~\bibnamefont {Cohen}}, \bibinfo {author} {\bibfnamefont
  {W.}~\bibnamefont {Dorland}}, \bibinfo {author} {\bibfnamefont
  {G.}~\bibnamefont {Hammett}}, \bibinfo {author} {\bibfnamefont
  {C.}~\bibnamefont {Kim}}, \bibinfo {author} {\bibfnamefont {J.}~\bibnamefont
  {Kinsey}}, \bibinfo {author} {\bibfnamefont {M.}~\bibnamefont
  {Kotschenreuther}}, \bibinfo {author} {\bibfnamefont {A.}~\bibnamefont
  {Kritz}}, \bibinfo {author} {\bibfnamefont {L.}~\bibnamefont {Lao}}, \bibinfo
  {author} {\bibfnamefont {J.}~\bibnamefont {Mandrekas}}, \bibinfo {author}
  {\bibfnamefont {W.}~\bibnamefont {Nevins}}, \bibinfo {author} {\bibfnamefont
  {S.}~\bibnamefont {Parker}}, \bibinfo {author} {\bibfnamefont
  {A.}~\bibnamefont {Redd}}, \bibinfo {author} {\bibfnamefont {D.}~\bibnamefont
  {Shumaker}}, \bibinfo {author} {\bibnamefont {R.Sydora}}, \ and\ \bibinfo
  {author} {\bibfnamefont {J.}~\bibnamefont {Weiland}},\ }\href@noop {}
  {\bibfield  {journal} {\bibinfo  {journal} {Phys. Plasmas}\ }\textbf
  {\bibinfo {volume} {7}},\ \bibinfo {pages} {969} (\bibinfo {year}
  {2000})}\BibitemShut {NoStop}%
\bibitem [{\citenamefont {Candy}\ and\ \citenamefont {Waltz}(2003)}]{gyro}%
  \BibitemOpen
  \bibfield  {author} {\bibinfo {author} {\bibfnamefont {J.}~\bibnamefont
  {Candy}}\ and\ \bibinfo {author} {\bibfnamefont {R.~E.}\ \bibnamefont
  {Waltz}},\ }\href@noop {} {\bibfield  {journal} {\bibinfo  {journal} {Journal
  of Computational Physics}\ }\textbf {\bibinfo {volume} {186}},\ \bibinfo
  {pages} {545 } (\bibinfo {year} {2003})}\BibitemShut {NoStop}%
\bibitem [{\citenamefont {Muhm}\ \emph {et~al.}(1992)\citenamefont {Muhm},
  \citenamefont {Pukhov}, \citenamefont {Spatschek},\ and\ \citenamefont
  {Tsytovich}}]{wk_derivation}%
  \BibitemOpen
  \bibfield  {author} {\bibinfo {author} {\bibfnamefont {A.}~\bibnamefont
  {Muhm}}, \bibinfo {author} {\bibfnamefont {A.}~\bibnamefont {Pukhov}},
  \bibinfo {author} {\bibfnamefont {K.}~\bibnamefont {Spatschek}}, \ and\
  \bibinfo {author} {\bibfnamefont {V.}~\bibnamefont {Tsytovich}},\ }\href@noop
  {} {\bibfield  {journal} {\bibinfo  {journal} {Phys. Fluids B}\ }\textbf
  {\bibinfo {volume} {4}} (\bibinfo {year} {1992})}\BibitemShut {NoStop}%
\bibitem [{\citenamefont {Toda}\ \emph {et~al.}(2006)\citenamefont {Toda},
  \citenamefont {Itoh}, \citenamefont {Hallatschek}, \citenamefont {Yagi},\
  and\ \citenamefont {Itoh}}]{todaItoh_zfscale}%
  \BibitemOpen
  \bibfield  {author} {\bibinfo {author} {\bibfnamefont {S.}~\bibnamefont
  {Toda}}, \bibinfo {author} {\bibfnamefont {K.}~\bibnamefont {Itoh}}, \bibinfo
  {author} {\bibfnamefont {K.}~\bibnamefont {Hallatschek}}, \bibinfo {author}
  {\bibfnamefont {M.}~\bibnamefont {Yagi}}, \ and\ \bibinfo {author}
  {\bibfnamefont {S.-I.}\ \bibnamefont {Itoh}},\ }\href@noop {} {\bibfield
  {journal} {\bibinfo  {journal} {J. Phys. Soc. Japan}\ }\textbf {\bibinfo
  {volume} {75}},\ \bibinfo {pages} {10} (\bibinfo {year} {2006})}\BibitemShut
  {NoStop}%
\bibitem [{\citenamefont {Smolyakov}\ \emph {et~al.}(2000)\citenamefont
  {Smolyakov}, \citenamefont {Diamond},\ and\ \citenamefont
  {Malkov}}]{smolyakov_wk}%
  \BibitemOpen
  \bibfield  {author} {\bibinfo {author} {\bibfnamefont {A.}~\bibnamefont
  {Smolyakov}}, \bibinfo {author} {\bibfnamefont {P.}~\bibnamefont {Diamond}},
  \ and\ \bibinfo {author} {\bibfnamefont {M.}~\bibnamefont {Malkov}},\
  }\href@noop {} {\bibfield  {journal} {\bibinfo  {journal} {PRL}\ }\textbf
  {\bibinfo {volume} {84}},\ \bibinfo {pages} {3} (\bibinfo {year}
  {2000})}\BibitemShut {NoStop}%
\bibitem [{\citenamefont {Dyachenko}\ \emph {et~al.}(1992)\citenamefont
  {Dyachenko}, \citenamefont {Nazarenko},\ and\ \citenamefont
  {Zakharov}}]{Dyachenko_wkintro}%
  \BibitemOpen
  \bibfield  {author} {\bibinfo {author} {\bibfnamefont {A.}~\bibnamefont
  {Dyachenko}}, \bibinfo {author} {\bibfnamefont {S.}~\bibnamefont
  {Nazarenko}}, \ and\ \bibinfo {author} {\bibfnamefont {V.}~\bibnamefont
  {Zakharov}},\ }\href@noop {} {\bibfield  {journal} {\bibinfo  {journal}
  {Phys. Letters A}\ }\textbf {\bibinfo {volume} {165}},\ \bibinfo {pages}
  {330} (\bibinfo {year} {1992})}\BibitemShut {NoStop}%
\bibitem [{\citenamefont {Lebedev}\ \emph {et~al.}(1995)\citenamefont
  {Lebedev}, \citenamefont {Diamond}, \citenamefont {Shapiro},\ and\
  \citenamefont {Soloview}}]{LebedevDiamond_wkintro}%
  \BibitemOpen
  \bibfield  {author} {\bibinfo {author} {\bibfnamefont {V.}~\bibnamefont
  {Lebedev}}, \bibinfo {author} {\bibfnamefont {P.}~\bibnamefont {Diamond}},
  \bibinfo {author} {\bibfnamefont {V.}~\bibnamefont {Shapiro}}, \ and\
  \bibinfo {author} {\bibfnamefont {G.}~\bibnamefont {Soloview}},\ }\href@noop
  {} {\bibfield  {journal} {\bibinfo  {journal} {Phys. Plasmas}\ }\textbf
  {\bibinfo {volume} {2}},\ \bibinfo {pages} {4421} (\bibinfo {year}
  {1995})}\BibitemShut {NoStop}%
\end{thebibliography}%
\end{document}